# Improved Frequency Ratio Measurement with $^{87}$Sr and $^{171}$Yb Optical Lattice Clocks at NMIJ

Yusuke Hisai[1, 2], Daisuke Akamatsu[1], Takumi Kobayashi[1], Kazumoto Hosaka[1], Hajime Inaba[1], Feng-Lei Hong[2], and Masami Yasuda[1]

[1] National Metrology Institute of Japan (NMIJ), National Institute of Advanced Industrial Science and Technology (AIST), 1-1-1 Umezono, Tsukuba, Ibaraki, 305-8563, Japan
[2] Department of Physics, Graduate School of Engineering Science, Yokohama National University, 79-5 Tokiwadai, Hodogaya-ku, Yokohama 240-8501, Japan

E-mail: d-akamatsu@aist.go.jp, takumi-kobayashi@aist.go.jp

## Abstract

We report improved frequency ratio measurement with $^{87}$Sr and $^{171}$Yb optical lattice clocks at the National Metrology Institute of Japan (NMIJ). The $^{87}$Sr optical lattice clock is enhanced with several major modifications and is re-evaluated with a reduced uncertainty of $1.1 \times 10^{-16}$. We employed a $^{171}$Yb optical lattice clock with an uncertainty of $4 \times 10^{-16}$ that was developed for contributing to International Atomic Time (TAI). The measurement result is $\nu_{\text{Yb}}/\nu_{\text{Sr}} = 1.207\,507\,039\,343\,338\,58(49)_{\text{sys}}(6)_{\text{stat}}$ with a fractional uncertainty of $4.1 \times 10^{-16}$, which is 3.4 times smaller than our previous measurement result.

Keywords: optical lattice clock, frequency metrology, frequency ratio, frequency measurement, SI second

## 1. Introduction

The second is the most precisely realized of all the fundamental units and is currently defined based on the microwave transition of a Cs atom. An atomic clock is an instrument that realizes the frequency of an atomic transition. The best Cs fountain clocks realize the current definition of the second at the accuracy of a few times $10^{-16}$ [1, 2].

Single-ion clocks and optical lattice clocks (OLCs), which are based on the optical transitions of atoms, have already achieved uncertainties at the $10^{-18} \sim 10^{-19}$ level [3-10]. Long-distance fibre links that allow us to compare optical clocks have been demonstrated [11]. Transportable OLCs have been developed [12] and used to compare optical clocks developed in different institutes [13] and to test general relativity [14]. SYRTE, NICT, NIST, INRIM and NMIJ have started to contribute to International Atomic Time (TAI) by using an optical clock as a second representation of the second [15]. These achievements are driving the discussion of the redefinition of the second using optical clocks.

By using two optical clocks with different atomic species, we can measure the ratio of the transition frequencies at the level of the clock uncertainties, which could be much smaller than that of the Cs fountain clock used for absolute frequency measurement. We can therefore validate the performance of the optical clocks beyond the accuracy of the Cs fountain clock by checking the consistency of independent frequency ratio measurements. Furthermore, checking the following identical equation with the frequency ratio results obtained in different laboratories can provide a stricter check of the clocks:

$$\frac{\nu_a}{\nu_b} \times \frac{\nu_b}{\nu_c} \times \frac{\nu_c}{\nu_a} = 1,$$

where $\nu_i$ ($i=a,b,c$) are the clock frequencies. Therefore, the precise measurement of the clock frequency ratio is listed as one of the milestones on the road to redefining the second [16]. Recently, the identical equation with Sr, Yb and Hg clock frequencies have been confirmed with an uncertainty of $1.3 \times 10^{-16}$ [17].

$^{87}$Sr and $^{171}$Yb OLCs are the most widely investigated of all the optical clocks. In 2014, we compared the frequencies of two OLCs, and determined the frequency ratio at a fractional uncertainty of $1.4 \times 10^{-15}$ [18, 19]. Ratio measurements with cryogenic $^{87}$Sr and $^{171}$Yb OLCs have been conducted by the RIKEN group [20, 21] with the smallest uncertainty of $4.6 \times 10^{-17}$. The frequency ratio has also been determined via a satellite link [22] and a transportable OLC [13]. We also reported the frequency ratio measured by the Sr-Yb dual-mode operation of an OLC [23]. Some of these ratio measurement results have been reported to the Consultative Committee for Time and Frequency (CCTF) under the International Committee for Weights and Measures (CIPM) as part of the discussion concerning the redefinition of the second.

In this paper, we report our second frequency ratio measurement result obtained with two independent $^{87}$Sr and $^{171}$Yb OLCs. We employed an improved Sr OLC, and a new Yb OLC that we built mainly to contribute to TAI [24]. We performed a re-evaluation of the systematic effects of our Sr OLC. The frequency ratio was determined as $1.207\,507\,039\,343\,338\,58(49)_{sys}(6)_{stat}$ with a fractional uncertainty of $4.1 \times 10^{-16}$. The measurement result will contribute to the discussion of the Yb/Sr frequency ratio at the next CCTF meeting.

## 2. Experimental method

An overview of our experimental setup is shown in Fig. 1. The $^{87}$Sr and $^{171}$Yb OLCs are situated on the same floor. We employ a linewidth transfer method to realize the clock lasers for probing $^{87}$Sr (698 nm) and $^{171}$Yb (578 nm) clock transitions using a fast controllable Er:fibre comb [25]. The fibre comb is phase-locked to an ultranarrow linewidth Nd:YAG laser at 1064 nm (master laser) stabilized to an ultrastable cavity. The fibre comb is equipped with a relocking system, which automatically recovers the phase-locking if the lock should fail [26]. The 698-nm clock laser for $^{87}$Sr is obtained directly from an ECDL. The 578-nm clock laser for interrogating the $^{171}$Yb clock transition is realized by the second harmonic generation of an external cavity diode laser (ECDL) at 1156 nm using a periodically poled lithium niobate (PPLN) waveguide. By phase-locking these lasers to the fibre comb, the linewidth of the master laser is transferred to these lasers [27].

Ti:sapphire lasers (M squared, SolsTiS) are employed to form vertically oriented 1-D optical lattices for $^{87}$Sr and $^{171}$Yb atoms. The optical lattices are formed by overlapping retroreflected beams on incoming beams using curved mirrors. Both lattice lasers are frequency-stabilized to other fibre combs referenced to the Coordinated Universal Time of NMIJ (UTC(NMIJ)) (not shown in Fig. 1) [28, 29]. The frequency fluctuation of the lattice lasers is less than 100 kHz during the experiment. The power of the lattice lasers is controlled using acousto-optic modulators (AOM1 and AOM2 in Fig. 1), so that the trap depths are stabilized at $46E_r^{Sr}$ for $^{87}$Sr and $450E_r^{Yb}$ for $^{171}$Yb, during spectroscopy. Here $E_r^{Sr(Yb)}$ is the photon recoil energy of the lattice laser for $^{87}$Sr ($^{171}$Yb).

The Sr OLC used in the present experiment is the same as that used in the previous absolute frequency [30, 31] and frequency ratio [18] measurements at NMIJ. In the present experiment, we made several modifications to the $^{87}$Sr OLC: 1) We installed a Doppler cancellation system to suppress the Doppler shift arising from the relative motion of the atoms trapped in the lattice against the clock laser [32]. As regards the $^{87}$Sr OLC operation, there is a mechanical shutter installed to shield the blackbody radiation (BBR) from the oven and an anti-Helmholtz coil for magneto-optical trapping (MOT) on the optical table, and they are quickly switched on and off.

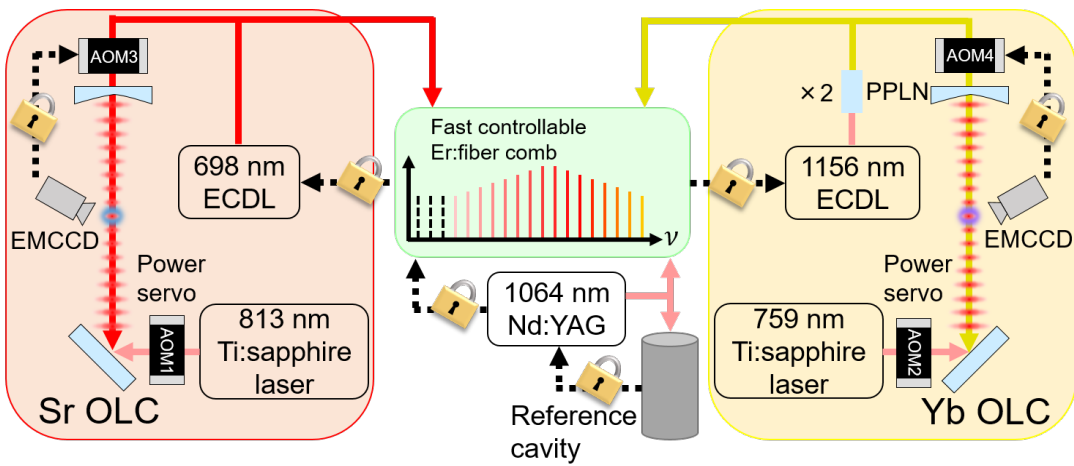

**Figure 1.** Schematic view of our $^{87}$Sr and $^{171}$Yb optical lattice clocks. AOM: acousto-optic modulator, EMCCD: electron multiplying charge-coupled device, ECDL: external cavity diode laser, PPLN: periodically poled lithium niobite waveguide.

Any possible mechanical vibrations, as mentioned above, are cancelled out by the Doppler cancellation system.
2) To evaluate the BBR shift, eight Pt1000 temperature sensors are newly attached to the vacuum chamber where the $^{87}$Sr atoms are trapped.
3) To pulse the clock laser, we frequency-shift the driving RF frequency of AOM3, instead of switching the RF power. This reduces the thermally induced AOM chirping effect.
The technical details and the result of the uncertainty evaluation of the $^{171}$Yb OLC have been reported elsewhere [24, 33].

Here, we give a simple description of the operation of the OLCs. $^{87}$Sr ($^{171}$Yb) atoms from an atomic oven are decelerated by Zeeman slowing using the $^1S_0$-$^1P_1$ transition. The slowed atoms are precooled with the MOT operating on the same transition (1st-stage MOT). After loading, the atoms are transferred to a second narrow-linewidth MOT with the $^1S_0$-$^3P_1$ transition (2nd-stage MOT). The atoms are then loaded in a linearly polarized one-dimensional optical lattice. The atoms are spin-polarized to a single magnetic sublevel of the clock transition $m_F = \pm 9/2$ for $^{87}$Sr ($m_F = \pm 1/2$ for $^{171}$Yb), and then two $\pi$ transitions are interrogated alternately. After the interrogation, the atoms remaining in the ground state are detected by applying a light resonant with the $^1S_0$-$^1P_1$ transition (imaging light). The fluorescence is detected by an electron multiplying charge-coupled device (EMCCD) camera. After this first step detection, a repumping laser repumps the atoms in the $^3P_0$ state to the ground state. The imaging light is employed again to detect the florescence. The excitation ratio of the clock transition is determined from these steps and used for clock laser frequency stabilization. Detailed descriptions of the setup and operation of the OLCs can be found in [30, 33].

## 3. Improved uncertainty evaluation of $^{87}$Sr OLC

The correction and uncertainty of the systematic effects of the $^{87}$Sr OLC after the re-evaluation in the present experiment are given in Table 1.

### 3.1 Lattice light shift

The intense optical lattice laser generally perturbs the clock transition frequency of the atoms. Katori proposed a "magic" wavelength $v_{\text{magic}}^{\text{E1}}$ where the laser induces the same amount of the frequency shift which originated from the electric dipole (E1) polarizability in the excited and ground states, resulting in cancellation of the shift in the clock transition frequency [34]. The higher order light shifts due to the electric-quadrupole (E2) and magnetic dipole (M1) polarizabilities and the hyperpolarizability are analysed in detail in [35].

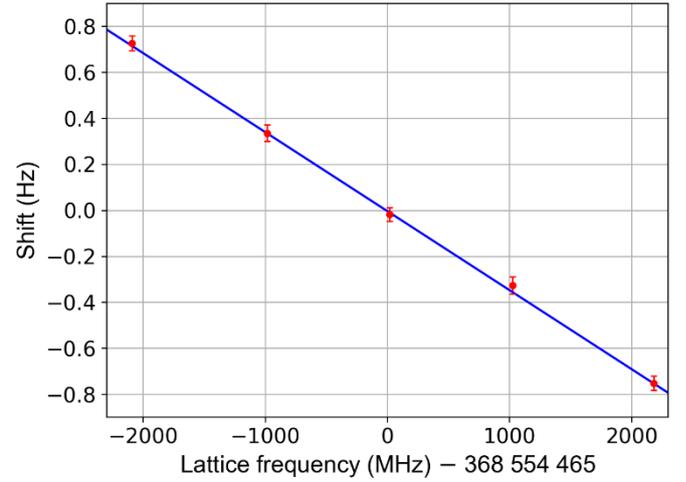

**Figure 2.** Lattice light shift measured with the interleave scheme. The trap depth was alternated between $51E_r^{\text{Sr}}$ and $23E_r^{\text{Sr}}$. The solid blue line represents a linear fit to the measured data.

To determine the magic wavelength $v_{\text{magic}}^{\text{E1}}$, we employed an interleaved scheme in which the frequency shift is measured by alternating the lattice depth between $51E_r^{\text{Sr}}$ and $23E_r^{Sr}$. The measured frequency shifts in five different lattice frequencies are plotted and fitted with a linear function in Fig. 2. We found the magic wavelength to be 368 554 463 (43) MHz. The determined magic wavelength has a smaller uncertainty than that obtained in our previous measurement [30]. The light shift at the operating lattice frequency including the higher order effects is evaluated to be $-2.4(5.7) \times 10^{-17}$ using the E2-M1 polarizabilities and the hyperpolarizability given in [35].

**Table 1.** The fractional corrections and uncertainties of the Sr and Yb optical lattice clocks.

|  | $^{87}$Sr clock | | $^{171}$Yb clock | |
|---|---|---|---|---|
|  | Correction | Uncertainty | Correction | Uncertainty |
| Systematic effect | ($\times 10^{-17}$) | | ($\times 10^{-17}$) | |
| Lattice light shift | 2.4 | 5.7 | -3.4 | 33.1 |
| BBR shift | 519.2 | 5.1 | 263.8 | 20.8 |
| Density shift | -4.5 | 4.8 | 3.8 | 3.7 |
| Quadratic Zeeman effect | 35.2 | <0.1 | 5.2 | 0.3 |
| DC Stark shift | 0 | 1.7 | 0 | <0.1 |
| Probe light shift | 2.0 | 1.1 | -0.4 | 0.2 |
| Servo error | 3.9 | 4.8 | 1.9 | 4.6 |
| Line pulling | 0 | 2 | 0 | 1 |
| AOM chirp | 0 | 1 | 0 | 1 |
| **Total** | **558.2** | **10.7** | **270.9** | **39.5** |
|  | Yb/Sr ratio | | | |
|  | Correction | Uncertainty | | |
| **Systematic effect** | **-287.3** | **40.9** | | |

### 3.2 Blackbody radiation shift

The BBR from the vacuum chamber induces the frequency shift of the clock transition, which is the dominant shift in our experiment. During clock operation, 8 sensors measure the temperature of the vacuum chamber every 10 seconds with the 4-point sensing method. In terms of the probability distribution of the effective temperature seen by the atoms at time $\tau$, we assume a rectangular distribution between the highest and lowest temperatures ($T_{max}(\tau)$ and $T_{min}(\tau)$) of the chamber. This distribution gives an average temperature of $T_{ave}(\tau) = (T_{max}(\tau) + T_{min}(\tau))/2$ with an uncertainty of $\Delta T_{ave}(\tau) = (T_{max}(\tau) - T_{min}(\tau))/\sqrt{12}$ [36]. We evaluated the BBR shift based on the formula

$$\Delta \nu_{BBR}(T_{ave}(\tau)) = \nu_{stat}\left(\frac{T_{ave}(\tau)}{T_0}\right)^4 + \nu_{dyn}\left(\frac{T_{ave}(\tau)}{T_0}\right)^6,$$

where $\nu_{stat}$ = -2.13023(6) Hz [37], $\nu_{dyn}$ = -0.1487(7) Hz [6] and $T_0$ = 300 K. The BBR shift is determined by averaging the shift throughout the measurement. The uncertainty of the instantaneous $\Delta\nu_{BBR}(\tau)$ is calculated using the uncertainty of temperature, $\Delta T_{ave}(\tau)$, and the maximum value of the calculated uncertainty is conservatively used as the evaluated uncertainty of the BBR shift in the present experiment. The improvement in the homogeneity of the temperature and the temperature measurement means that the uncertainty of the BBR shift obtained in the present experiment is smaller by a factor of more than 3 compared with that obtained in our previous work [30].

We assumed that the BBR emitted from the oven does not contribute to the systematic shift since the mechanical shutter in front of the oven is closed during spectroscopy.

### 3.3 Density shift

In the $^{87}$Sr OLC, a frequency shift caused by $s$-wave collisions is suppressed by spin-polarizing the atoms into $m_F = \pm 9/2$ hyperfine states based on the Pauli exclusion principle. However, imperfect spin polarization or inhomogeneous excitation could cause a collisional shift [38]. We investigate the collisional shift by measuring the density shift with an interleave scheme. This is designed to measure the frequency shift by alternating the atom number in the lattice (proportional to the atomic density), which is realized by changing the power of the cooling laser for the 1st-stage MOT.

The atom number in the lattice is measured using the fluorescence signal detected with the EMCCD. The frequency difference for the different atom numbers ($N_a$, $N_b$) corresponds to the frequency shift in the experiment induced by the variation of the atom numbers $\Delta N = N_a - N_b$.

The frequency shift as a function of the atomic density is shown in Fig. 3. The solid and broken lines represent the linear fit of the measured data and its uncertainty, respectively. The correction of the density shift is estimated to be $4.5(4.8) \times 10^{-17}$ under our experimental conditions (indicated by a triangle in Fig. 3). The horizontal error bars indicate the uncertainty of the atomic density caused by the intensity fluctuation of the imaging light during the measurement. The uncertainty of the density shift obtained in the present experiment is more than a factor of 4 smaller than that obtained in our previous work [30].

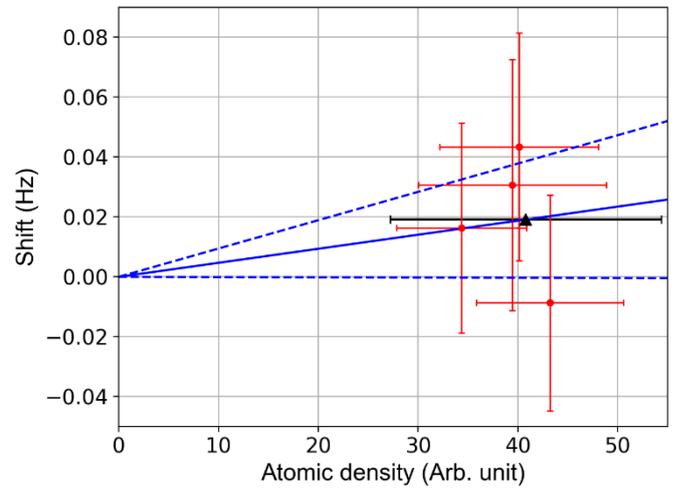

**Figure 3.** Density shift as a function of atomic density. The solid (broken) line represents a linear fit (its uncertainty) to the measured data. The triangle indicates the experimental condition of the frequency ratio measurement.

### 3.4 Zeeman shift

To determine the unperturbed frequency of the clock transition, we intentionally applied a magnetic field (|**B**| ~ 160 µT) to lift the degeneracy of the Zeeman sublevels. The 1st order Zeeman effect splits the clock transition into $m_F = \pm 9/2$ sub-states by $\Delta\nu_{splitting}$ ~ 780 Hz. Although the 1st order Zeeman effect can be cancelled by averaging the frequencies of the $\pi$ transitions of $m_F = \pm 9/2$, the quadratic Zeeman shift remains. From the coefficient obtained in [10], we determined the quadratic Zeeman shift to be $-35.2 \times 10^{-17}$ for an applied magnetic field with an uncertainty of less than $1 \times 10^{-18}$, which is mainly limited by that of the coefficient.

### 3.5 Other shifts

The stray electric field on the viewports of the vacuum chamber, which are 85 mm from the atoms, causes a DC stark shift. After the ratio measurement, we measured the DC electric potential on the viewports with an electrostatic sensor (Keyence SK-050) and found it to be less than 30 V (measurement uncertainty 10 V). With this result, we evaluated the electric field at the atoms with a finite element method and determined the DC Stark shift to be $0(1.7) \times 10^{-17}$ using the reported static polarizability [37].

We evaluated the light shift caused by the probe laser employed in Rabi spectroscopy using the value reported in [6]. The laser intensity is inversely proportional to the square of the interrogation time for the Rabi π pulse. From this fact, we determined the probe light shift to be $-2.0(1.1) \times 10^{-17}$ for our pulse duration of 40 ms.

We also evaluated the frequency shift caused by the servo error and found it to be $-3.9(4.8) \times 10^{-17}$. The uncertainty of the line pulling induced by the imperfect spin-polarization is conservatively estimated to be $2 \times 10^{-17}$. As mentioned in section 2, we suppressed the effect of the frequency shift caused by AOM chirping in the clock pulse. We conservatively estimated the uncertainty resulting from AOM chirping to be $1 \times 10^{-17}$ as in [33].

## 4. Frequency ratio measurement of Yb/Sr

The frequency ratio of the $^{171}$Yb and $^{87}$Sr clock transitions was measured during the continuous operation of an Yb OLC since September 2019 [24]. The Sr OLC was operated for 6 days with a total ratio measurement time of about 70000 s. A typical Allan standard deviation of the frequency ratio measurement in one day is shown in Fig. 4. The instability follows $1.2 \times 10^{-14} \, (\tau/s)^{-1/2}$ after 30 s. The frequency ratio measurement result for each day is plotted in Fig. 5. The black and red uncertainty bars represent the statistical and total uncertainties, respectively. The systematic corrections and the uncertainties of the Yb/Sr ratio are shown in Table 1. The gravitational shift caused by the height difference of the atoms (1 cm) is neglected. We determine the $^{171}$Yb/$^{87}$Sr ratio to be $\nu_{Yb}/\nu_{Sr} = 1.207\,507\,039\,343\,338\,58(49)_{sys}(6)_{stat}$. The total fractional uncertainty is $4.1 \times 10^{-16}$, which is 3.4 times smaller than the result we obtained in 2014 [19]. The total uncertainty of the optical frequency measurement is mainly limited by the systematic uncertainty of the $^{171}$Yb OLC as shown in Table 1.

Figure 6 shows the present result together with previous frequency ratio measurement results obtained at different institutes. While the NMIJ2014 [19], RIKEN2015 [20], RIKEN2016 [21] and the present result were obtained by local comparisons of OLCs, KRISS/NICT2018 [22] was obtained using a satellite link. INRIM/PTB2018 [13] and NMIJ2018 [23] were obtained by using a transportable OLC and the dual-mode operation of an OLC, respectively. INRIM/(NICT,SYRTE)2020 [39] and NMIJ/NICT2020 [24] were obtained utilizing a link to TAI. Our present result agrees with the ratio calculated using the recommended frequencies of Sr and Yb by the CIPM [16], within the combined uncertainty. Our present result also agrees with results reported by several institutes within the measurement uncertainties, but only agrees with the RIKEN2016 result (which has the smallest measurement uncertainty) within the 2-Sigma uncertainty of our result.

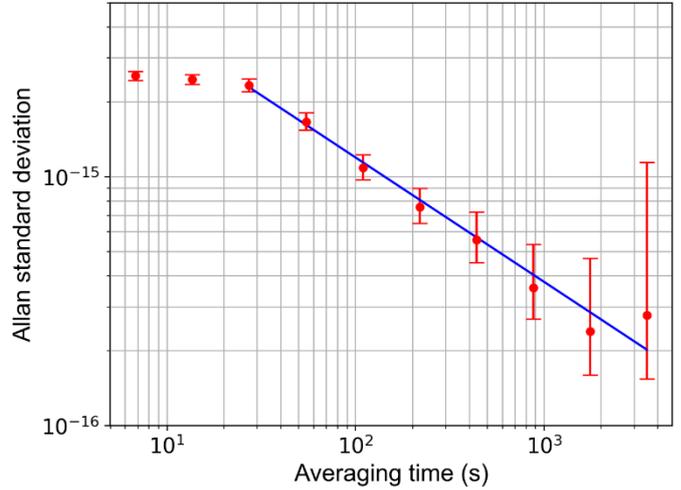

**Figure 4.** Fractional instability of the measured $^{171}$Yb/$^{87}$Sr frequency ratio.

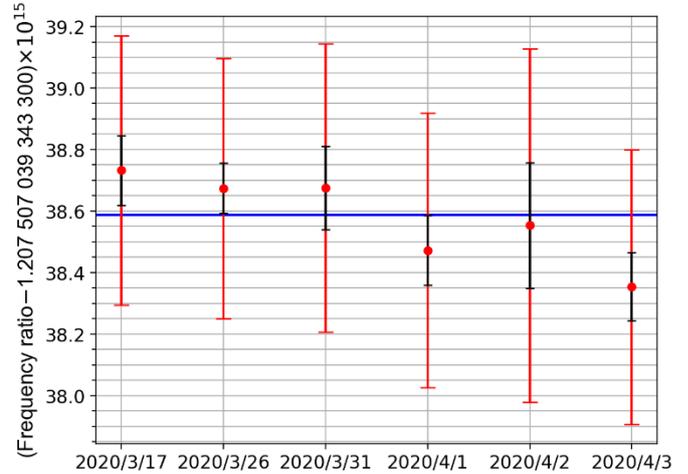

**Figure 5.** $^{171}$Yb/$^{87}$Sr frequency ratio measurements over 6 days. The black and red uncertainty bars represent statistical and total uncertainties, respectively.

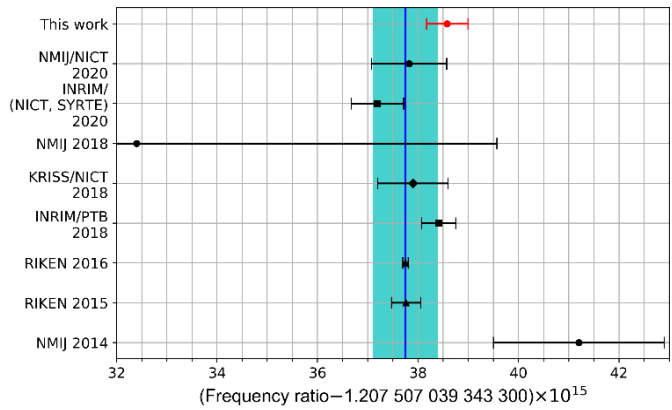

**Figure 6.** A history of Yb/Sr frequency ratio measurements. The blue line and shaded region represent the frequency ratio calculated using the recommended frequencies of $^{87}$Sr and $^{171}$Yb by the CIPM and its uncertainty, respectively.

## 5. Discussion and conclusion

The main contributors to the uncertainty of our Sr OLC are the systematic effects of the BBR shift, the lattice light shift, the density shift and the servo error. With further precise uncertainty evaluations, the uncertainty contributions from the lattice light shift, density shift and servo error could be further reduced. However, the uncertainty from the BBR shift has already reached its limitation at around $5 \times 10^{-17}$ unless we introduce a cryogenic environment or BBR shield for the interaction area of the clock [5, 8, 10]. For the Yb OLC, the main contributors to the uncertainty are the systematic effects of the BBR and the lattice light shifts. The larger uncertainties in the BBR and the lattice light shifts of the Yb OLC compared with that of the Sr OLC originated from the heated window inside the vacuum and the larger trap potential, respectively [24]. These experimental conditions are necessary for the long-term stable operation of the $^{171}$Yb OLC, and the associated uncertainties may be improved by further careful evaluations in the future.

Within the framework of the possible future redefinition of the second using optical clocks [16], it is important to achieve TAI calibration based on regular reports from metrological institutes where optical clocks are in operation. We have realized the nearly continuous operation of our $^{171}$Yb OLC for half a year and reported the evaluated frequency of TAI to BIPM [24]. The uncertainty of the frequency reported monthly to BIPM is limited by either the link uncertainty of the UTC(NMIJ) to TAI, or the optical clock uncertainty, or the link uncertainty between the optical clock and the UTC(NMIJ). The link uncertainty of the UTC(NMIJ) to TAI in one month is approximately $2.0 \times 10^{-16}$. The uncertainty of our $^{87}$Sr OLC is $1.1 \times 10^{-16}$. The link uncertainty between the optical clock and the UTC(NMIJ) is limited by the uptime of the optical clock. To reach an uncertainty for the link between the $^{87}$Sr OLC and the UTC(NMIJ) of smaller than $2.0 \times 10^{-16}$, we need to operate the Sr OLC with an uptime of $> 50\%$ every day. This may require the updating of the laser systems in the Sr OLC. In the application of the contribution to TAI, it is important to keep the system simple and robust for long-term operation, rather than to introduce a cryogenic environment or a BBR shield into the interaction area for further uncertainty reduction.

Any report of an optical clock for TAI calibration is usually reviewed by a working group of experts under the CIPM. During the review process, the experts will investigate any possible doubt related to the measurement. If we operate two optical clocks for TAI calibration and measure their frequency ratio at the same time, the reliability of the measurement can be greatly improved. This will result in a triangular comparison between the two optical clocks and the SI second. Since the instability of the ratio measurement reaches $1 \times 10^{-16}$ at an averaging time of about 10000 s (about 3 hours), the ratio measurement provides instant monitoring of the clock frequencies during a long-term measurement for TAI calibration. Furthermore, with more data for such triangular comparisons from different institutes containing different clocks, it will be possible to verify the consistency of optical clocks worldwide and also enhance the link between the optical clocks and the SI second, using the matrix method [40]. Such investigations should certainly contribute to discussions regarding the redefinition of the second.

In conclusion, we measured the frequency ratio of the $^{87}$Sr and $^{171}$Yb clock transitions with the OLCs at NMIJ. The frequency ratio was determined as 1.207 507 039 343 338 58(49)$_{sys}$(6)$_{stat}$ with a fractional uncertainty of $4.1 \times 10^{-16}$. Thanks to the improvement in the OLCs, the uncertainty is reduced by a factor of 3.4 compared with our previously reported value.

## Acknowledgements

This work was supported by the Japan Society for the Promotion of Science (JSPS) KAKENHI Grant Number 17H01151, and 17K14367, 18H03886, 19J10824, and JST-Mirai Program Grant Number JPMJMI18A1, Japan.

## References

[1] Guéna J, Abgrall M, Rovera D, Laurent P, Chupin B, Lours M, Santarelli G, Rosenbusch P, Tobar M E, Li R, Gibble K, Clairon A and Bize S 2012 Progress in atomic fountains at LNE-SYRTE *IEEE Trans. Ultrason. Ferroelectr. Freq. Control* **59** 391-409
[2] Weyers S, Gerginov V, Kazda M, Rahm J, Lipphardt B, Dobrev G and Gibble K 2018 Advances in the accuracy, stability, and

reliability of the PTB primary fountain clocks *Metrologia* **55** 789–805
[3] Chou C W, Hume D B, Koelemeij J C J, Wineland D J and Rosenband T 2010 Frequency comparison of two high-accuracy Al[+] optical clocks *Phys. Rev. Lett.* **104** 070802
[4] Bloom B J, Nicholson T L, Williams J R, Campbell S L, Bishof M, Zhang X, Zhang W, Bromley S L and Ye J 2014 An optical lattice clock with accuracy and stability at the $10^{-18}$ level *Nature* **506** 71–5
[5] Ushijima I, Takamoto M, Das M, Ohkubo T and Katori H 2015 Cryogenic optical lattice clocks *Nat. Photonics* **9** 185–9
[6] Nicholson T L, Campbell S L, Hutson R B, Marti G E, Bloom B J, McNally R L, Zhang W, Barrett M D, Safronova M S, Strouse G F, Tew W L and Ye J 2015 Systematic evaluation of an atomic clock at $2 \times 10^{-18}$ total uncertainty *Nat. Commun.* **6** 6896
[7] Huntemann N, Sanner C, Lipphardt B, Tamm C and Peik E 2016 Single-Ion Atomic Clock with $3 \times 10^{-18}$ Systematic Uncertainty *Phys. Rev. Lett.* **116** 063001
[8] McGrew W F, Zhang X, Fasano R J, Schäffer S A, Beloy K, Nicolodi D, Brown R C, Hinkley N, Milani G, Schioppo M, Yoon T H and Ludlow A D 2018 Atomic clock performance enabling geodesy below the centimetre level *Nature* **564** 87–90
[9] Brewer S M, Chen J S, Hankin A M, Clements E R, Chou C W, Wineland D J, Hume D B and Leibrandt D R 2019 [27]Al[+] Quantum-Logic Clock with a Systematic Uncertainty below $10^{-18}$ *Phys. Rev. Lett.* **123** 033201
[10] Bothwell T, Kedar D, Oelker E, Robinson J M, Bromley S L, Tew W L, Ye J and Kennedy C J 2019 JILA SrI optical lattice clock with uncertainty of $2.0 \times 10^{-18}$ *Metrologia* **56** 065004
[11] Lisdat C, Grosche G, Quintin N, Shi C, Raupach S M F, Grebing C, Nicolodi D, Stefani F, Al-Masoudi A, Dörscher S, Häfner S, Robyr J-L, Chiodo N, Bilicki S, Bookjans E, Koczwara A, Koke S, Kuhl A, Wiotte F, Meynadier F, Camisard E, Abgrall M, Lours M, Legero T, Schnatz H, Sterr U, Denker H, Chardonnet C, Coq Y Le, Santarelli G, Amy-Klein A, Targat R Le, Lodewyck J, Lopez O and Pottie P-E 2016 A clock network for geodesy and fundamental science *Nat. Commun.* **7** 12443
[12] Koller S B, Grotti J, Vogt S, Al-Masoudi A, Dörscher S, Häfner S, Sterr U and Lisdat C 2017 Transportable Optical Lattice Clock with $7 \times 10^{-17}$ Uncertainty *Phys. Rev. Lett.* **118** 073601
[13] Grotti J, Koller S, Vogt S, Häfner S, Sterr U, Lisdat C, Denker H, Voigt C, Timmen L, Rolland A, Baynes F N, Margolis H S, Zampaolo M, Thoumany P, Pizzocaro M, Rauf B, Bregolin F, Tampellini A, Barbieri P, Zucco M, Costanzo G A, Clivati C, Levi F and Calonico D 2018 Geodesy and metrology with a transportable optical clock *Nat. Phys.* **14** 437–41
[14] Takamoto M, Ushijima I, Ohmae N, Yahagi T, Kokado K, Shinkai H and Katori H 2020 Test of general relativity by a pair of transportable optical lattice clocks *Nat. Photonics* **14** 411–5
[15] https://www.bipm.org/en/bipm-services/timescales/time-ftp/Circular-T.html
[16] Riehle F, Gill P, Arias F and Robertsson L 2018 The CIPM list of recommended frequency standard values: guidelines and procedures *Metrologia* **55** 188–200
[17] Ohmae N, Bregolin F, Nemitz N and Katori H 2020 Direct measurement of the frequency ratio for Hg and Yb optical lattice clocks and closure of the Hg/Yb/Sr loop *Opt. Express* **28** 15112–21
[18] Akamatsu D, Yasuda M, Inaba H, Hosaka K, Tanabe T, Onae A and Hong F-L 2014 Frequency ratio measurement of [171]Yb and [87]Sr optical lattice clocks *Opt. Express* **22** 7898–905
[19] Akamatsu D, Yasuda M, Inaba H, Hosaka K, Tanabe T, Onae A and Hong F-L 2014 Errata: Frequency ratio measurement of [171]Yb and [87]Sr optical lattice clocks *Opt. Express* **22** 32199
[20] Takamoto M, Ushijima I, Das M, Nemitz N, Ohkubo T, Yamanaka K, Ohmae N, Takano T, Akatsuka T, Yamaguchi A and Katori H 2015 Frequency ratios of Sr, Yb, and Hg based optical lattice clocks and their applications *Comptes Rendus Phys.* **16** 489–98
[21] Nemitz N, Ohkubo T, Takamoto M, Ushijima I, Das M, Ohmae N and Katori H 2016 Frequency ratio of Yb and Sr clocks with $5 \times 10^{-17}$ uncertainty at 150 seconds averaging time *Nat. Photonics* **10** 258–61
[22] Fujieda M, Yang S H, Gotoh T, Hwang S W, Hachisu H, Kim H, Lee Y K, Tabuchi R, Ido T, Lee W K, Heo M S, Park C Y, Yu D H and Petit G 2018 Advanced Satellite-Based Frequency Transfer at the $10^{-16}$ Level *IEEE Trans. Ultrason. Ferroelectr. Freq. Control* **65** 973–8
[23] Akamatsu D, Kobayashi T, Hisai Y, Tanabe T, Hosaka K, Yasuda M and Hong F-L 2018 Dual-mode operation of an optical lattice clock using strontium and ytterbium atoms *IEEE Trans. Ultrason. Ferroelectr. Freq. Control* **65** 1069–75
[24] Kobayashi T, Akamatsu D, Hosaka K, Hisai Y, Wada M, Inaba H, Suzuyama T, Hong F-L, and Yasuda M 2020 Demonstration of the nearly continuous operation of an [171]Yb optical lattice clock for half a year *Metrologia* (to be published)
[25] Nakajima Y, Inaba H, Hosaka K, Minoshima K, Onae A, Yasuda M, Kohno T, Kawato S, Kobayashi T, Katsuyama T and Hong F-L 2010 A multi-branch, fiber-based frequency comb with millihertz-level relative linewidths using an intra-cavity electro-optic modulator *Opt. Express* **18** 1667–76
[26] Kobayashi T, Akamatsu D, Hosaka K and Yasuda M 2019 A relocking scheme for optical phase locking using a digital circuit with an electrical delay line *Rev. Sci. Instrum.* **90** 103002
[27] Inaba H, Hosaka K, Yasuda M, Nakajima Y, Iwakuni K, Akamatsu D, Okubo S, Kohno T, Onae A and Hong F-L 2013 Spectroscopy of [171]Yb in an optical lattice based on laser linewidth transfer using a narrow linewidth frequency comb *Opt. Express* **21** 7891-7896
[28] Hisai Y, Ikeda K, Sakagami H, Horikiri T, Kobayashi T, Yoshii K and Hong F-L 2018 Evaluation of laser frequency offset locking using an electrical delay line *Appl. Opt.* **57** 5628–34
[29] Hisai Y, Akamatsu D, Kobayashi T, Okubo S, Inaba H, Hosaka K, Yasuda M and Hong F-L 2019 Development of 8-branch Er:fiber frequency comb for Sr and Yb optical lattice clocks *Opt. Express* **27** 6404–14

[30] Akamatsu D, Inaba H, Hosaka K, Yasuda M, Onae A, Suzuyama T, Amemiya M and Hong F-L 2014 Spectroscopy and frequency measurement of the $^{87}$Sr clock transition by laser linewidth transfer using an optical frequency comb *Appl. Phys. Express* **7** 012401

[31] Tanabe T, Akamatsu D, Kobayashi T, Takamizawa A, Yanagimachi S, Ikegami T, Suzuyama T, Inaba H, Okubo S, Yasuda M, Hong F-L, Onae A and Hosaka K 2015 Improved Frequency Measurement of the $^1S_0$-$^3P_0$ Clock Transition in $^{87}$Sr Using a Cs Fountain Clock as a Transfer Oscillator *J. Phys. Soc. Japan* **84** 115002

[32] Takamoto M, Takano T and Katori H 2011 Frequency comparison of optical lattice clocks beyond the Dick limit *Nat. Photonics* **5** 288–92

[33] Kobayashi T, Akamatsu D, Hisai Y, Tanabe T, Inaba H, Suzuyama T, Hong F-L, Hosaka K and Yasuda M 2018 Uncertainty Evaluation of an $^{171}$Yb Optical Lattice Clock at NMIJ *IEEE Trans. Ultrason. Ferroelectr. Freq. Control* **65** 2449–58

[34] Katori H, Takamoto M, Pal'chikov V G and Ovsiannikov V D 2003 Ultrastable optical clock with neutral atoms in an engineered light shift trap. *Phys. Rev. Lett.* **91** 173005

[35] Ushijima I, Takamoto M and Katori H 2018 Operational Magic Intensity for Sr Optical Lattice Clocks *Phys. Rev. Lett.* **121** 263202

[36] BIPM, IEC, IFCC, ILAC, ISO, IUPAC, IUPAP and OIML 2008 Evaluation of measurement data Guide to the Expression of Uncertainty in Measurement (Joint Committee for Guides in Metrology vol 100) (Geneve: International Organization for Standardization)

[37] Middelmann T, Falke S, Lisdat C and Sterr U 2012 High accuracy correction of blackbody radiation shift in an optical lattice clock *Phys. Rev. Lett.* **109** 263004

[38] Campbell G K, Boyd M M, Thomsen J W, Martin M J, Blatt S, Swallows M D, Nicholson T L, Fortier T, Oates C W, Diddams S A, Lemke N D, Naidon P, Julienne P, Ye J and Ludlow A D, 2009 Probing Interactions Between Ultracold Fermions *Science* **324** 360

[39] Pizzocaro M, Bregolin F, Barbieri P, Rauf B, Levi F and Calonico D 2020 Absolute frequency measurement of the $^1S_0$-$^3P_0$ transition of $^{171}$Yb with a link to international atomic time *Metrologia* **57** 035007

[40] Margolis H S and Gill P 2015 Least-squares analysis of clock frequency comparison data to deduce optimized frequency and frequency ratio values *Metrologia* **52** 628–34